\documentclass{webofc}
\usepackage[varg]{txfonts}   
\usepackage{graphicx}
\usepackage{hyperref}
\usepackage{xspace}
\usepackage{fontawesome5}
\usepackage{xcolor}

\newcommand{\runII}{\ensuremath{\mathrm{Run~2}}\xspace}
\newcommand{\runIII}{\ensuremath{\mathrm{Run~3}}\xspace}
\newcommand{\gaudi}{\ensuremath{\mbox{\sc Gaudi}}\xspace}
\newcommand{\gauss}{\ensuremath{\mbox{\sc Gauss}}\xspace}
\newcommand{\gaussino}{\ensuremath{\mbox{\sc Gaussino}}\xspace}

\newcommand{\pythia}{\ensuremath{\mbox{\sc Pythia8}}\xspace}
\newcommand{\evtgen}{\ensuremath{\mbox{\sc EvtGen}}\xspace}
\newcommand{\geant}{\ensuremath{\mbox{\sc Geant4}}\xspace}
\newcommand{\lamarr}{\ensuremath{\mbox{\sc Lamarr}}\xspace}
\newcommand{\flashsim}{\ensuremath{\mbox{\sc FlashSim}}\xspace}
\newcommand{\delphes}{\ensuremath{\mbox{\sc Delphes}}\xspace}
\newcommand{\LbLcmunu}{\ensuremath{\Lambda_b^0 \to \Lambda_c^+ \mu^- \bar \nu_\mu}\xspace}
\newcommand{\LcpKpi}{\ensuremath{\Lambda_c^+ \to p K^- \pi^+}\xspace}

\newcommand{\orcid}[1]{\href{https://orcid.org/#1}{\textcolor[HTML]{A6CE39}{\faOrcid}}}
\begin{document}
\title{The LHCb ultra-fast simulation option, Lamarr}
\subtitle{design and validation}




\author{
    \firstname{Lucio}~\lastname{Anderlini}\inst{1} \and
    \firstname{Matteo}~\lastname{Barbetti}\inst{1,2}\fnsep\thanks{
      \email{matteo.barbetti@cern.ch}
    } \and
    \firstname{Simone}~\lastname{Capelli}\inst{3,4} \and
    \firstname{Gloria}~\lastname{Corti}\inst{5} \and
    \firstname{Adam}~\lastname{Davis}\inst{6} \and
    \firstname{Denis}~\lastname{Derkach}\inst{7} \and
    \firstname{Nikita}~\lastname{Kazeev}\inst{7} \and
    \firstname{Artem}~\lastname{Maevskiy}\inst{7} \and
    \firstname{Maurizio}~\lastname{Martinelli}\inst{3,4} \and
    \firstname{Sergei}~\lastname{Mokonenko}\inst{7} \and
    \firstname{Benedetto~G.}~\lastname{Siddi}\inst{8} \and
    \firstname{Zehua}~\lastname{Xu}\inst{9}~on behalf of the LHCb Simulation Project
}

\institute{
  Istituto Nazionale di Fisica Nucleare (INFN),
  Sezione di Firenze, Italy \and
  Department of Information Engineering (DINFO),
  University of Firenze, Italy \and
  Istituto Nazionale di Fisica Nucleare (INFN),
  Sezione di Milano-Bicocca, Italy \and
  Department of Physics,
  University of Milano-Bicocca, Italy \and
  European Organization for Nuclear Research (CERN),
  Switzerland \and
  Department of Physics and Astronomy, 
  University of Manchester, United Kingdom \and
  Affiliated with an institute covered by 
  a cooperation agreement with CERN \and
  Department of Physics, 
  University of Ferrara, Italy \and
  Laboratoire de Physique de Clermont (LPC), 
  Université Clermont Auvergne, France
}

\abstract{
  Detailed detector simulation is the major consumer of CPU resources 
  at LHCb, having used more than 90\% of the total computing budget 
  during Run 2 of the Large Hadron Collider at CERN. As data is 
  collected by the upgraded LHCb detector during Run 3 of the LHC, 
  larger requests for simulated data samples are necessary, and will 
  far exceed the pledged resources of the experiment, even with existing
  fast simulation options. An evolution of technologies and techniques 
  to produce simulated samples is mandatory to meet the upcoming needs 
  of analysis to interpret signal versus background and measure 
  efficiencies. In this context, we propose \lamarr, a \gaudi-based 
  framework designed to offer the fastest solution for the 
  simulation of the LHCb detector.
  \lamarr consists of a pipeline of modules parameterizing both the 
  detector response and the reconstruction algorithms of the LHCb 
  experiment. Most of the parameterizations are made of Deep Generative 
  Models and Gradient Boosted Decision Trees trained on simulated 
  samples or alternatively, where possible, on real data. Embedding 
  \lamarr in the general LHCb \gauss Simulation framework allows combining 
  its execution with any of the available generators in a seamless way.
  \lamarr has been validated by comparing key reconstructed quantities 
  with Detailed Simulation. Good agreement of the simulated 
  distributions is obtained with two-order-of-magnitude speed-up of 
  the simulation phase.
}

\maketitle

\section{Introduction}
\label{sec:intro}

The LHCb experiment~\cite{LHCb:2008vvz} has been originally designed 
to study rare decays of particles containing $b$ and $c$ quarks 
produced at the Large Hadron Collider~(LHC). The LHCb detector is 
a single-arm forward spectrometer covering the pseudorapidity 
range of $2 < \eta < 5$, that includes a Tracking system and 
a Particle Identification (PID) system~\cite{LHCb:2014set}. The 
Tracking system provides high-precision measurements of the 
momentum~$p$ of charged particles and the position of primary 
vertices. Different types of charged hadrons are separated 
using the response of two ring-imaging Cherenkov~(RICH) detectors. 
Photons, electrons and hadrons are identified by the 
calorimeter system relying on an electromagnetic 
calorimeter~(ECAL) and a hadron calorimeter~(HCAL). 
Finally, a dedicated system named MUON identifies muons alternating 
layers of iron and multi-wire proportional chambers. The RICH, 
calorimeters and MUON detectors are part of the PID system.

Interpreting signal, rejecting background contributions and performing 
efficiency studies requires to have a full understanding of its 
data sample, from the high-energy collisions to the set of physics 
processes responsible for the detector high-level response. 
This kind of studies greatly benefits from the use of simulated 
samples. At LHCb, the simulation production mainly relies on 
the \gauss framework~\cite{Clemencic:2011zza} that implements 
the generation and simulation phases, and is based on the 
\gaudi processing framework~\cite{Barrand:2001ny}. 
The high-energy collisions and all the physics processes that produce 
the set of particles (e.g., muons, pions, kaons or protons) able 
to traverse the LHCb spectrometer are simulated during the 
\emph{generation phase} using software like 
\pythia~\cite{Sjostrand:2007gs} and \evtgen~\cite{Lange:2001uf}.
The radiation-matter interactions between the detector materials
and the traversing particles are reproduced during the 
\emph{simulation phase} that aims to compute the energy deposited 
in the active volumes and relies on the \geant toolkit~\cite{Allison:2006ve}.
Then, a separate application converts the energy deposits into
raw data compatible with the real one collected by LHCb.

The simulation of all the physics events occurring within the detector
is the major consumer of CPU resources at LHCb, having used more 
than 90\% of the total computing budget during LHC \runII. The
upgraded version of the experiment is designed to collect 
one-order-of-magnitude larger data samples during \runIII. Meeting 
the upcoming and future requests for simulated samples is not 
sustainable relying only on the traditional \emph{detailed simulation}. 
For this reason, the LHCb Collaboration is spending great efforts 
in modernizing the simulation software stack through the novel 
experiment-independent framework \gaussino\footnote{Visit 
\url{https://gaussino.docs.cern.ch} 
for additional details.}~\cite{Mazurek:2021abc, Mazurek:2022tlu} 
on which a newer version of \gauss will be built on, and in 
developing faster simulation options, some of which also powered by 
machine learning algorithms~\cite{Chekalina:2018hxi,
Maevskiy:2019vwj, Anderlini:2022ofl, Barbetti:2023bvi}.

\section{Fast simulation VS. ultra-fast simulation}
\label{sec:fast-sim}

\begin{figure}[h!]
  \begin{minipage}{0.45\textwidth}
    \centering
    \includegraphics[width=\textwidth]{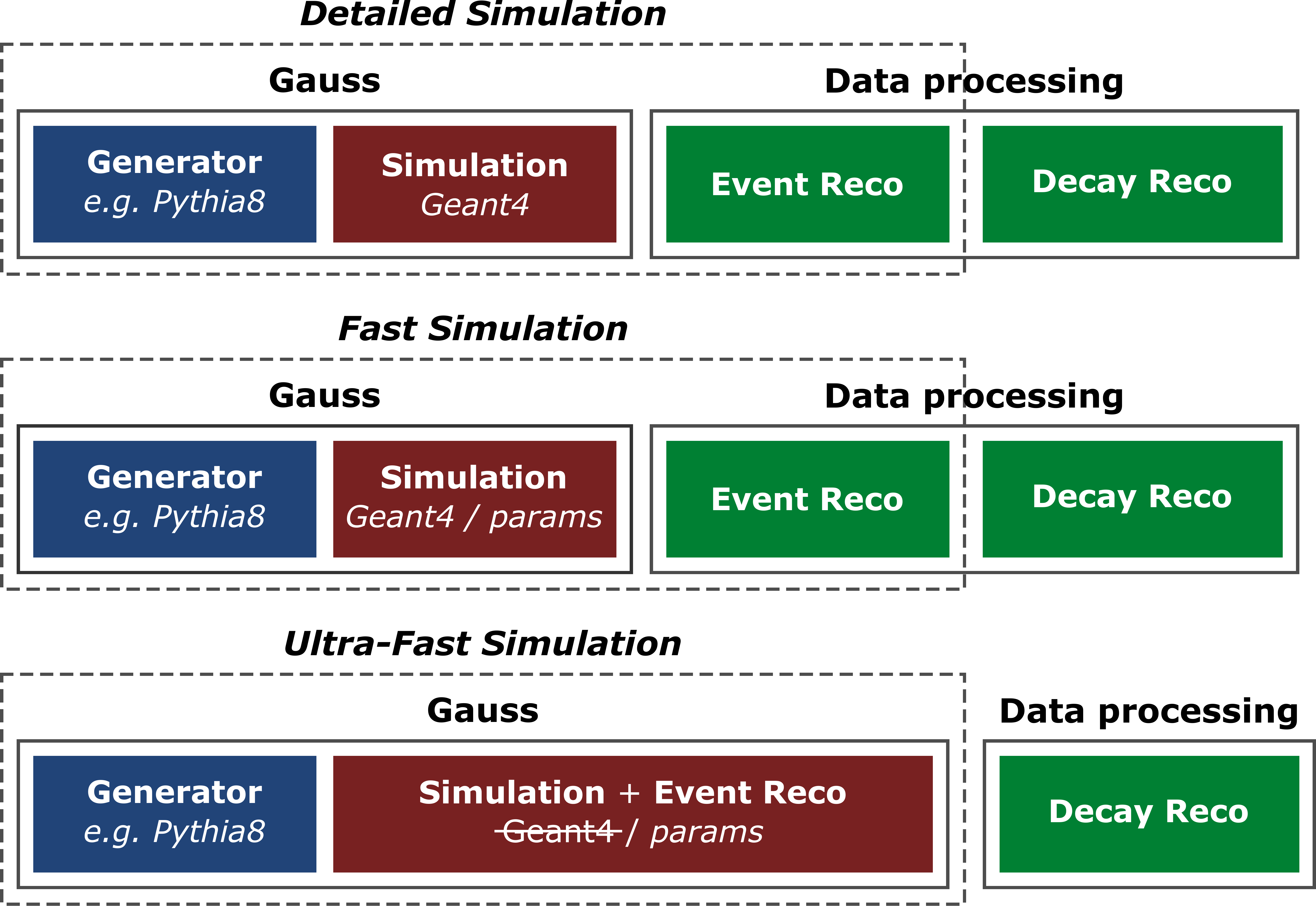}
  \end{minipage}
  \hfill
  \begin{minipage}{0.52\textwidth}
    \caption{
      \label{fig:sim-proc-flow}
      Schematic representation of the data processing flow in the
      \emph{detailed} (top), \emph{fast} (center) and 
      \emph{ultra-fast} (bottom) simulation paradigms.
    }
  \end{minipage}
\end{figure}

Simulating all the physics processes of interest for LHCb is 
extremely expensive in terms of computing resources, especially 
the \geant-based step that is the major CPU-consumer. 
Speeding up the computation of the energy deposits or, more generally, 
the detector response is mandatory to satisfy the demand for simulations
expected for \runIII and those that will follow. Actually, this is a
shared problem across the High Energy Physics (HEP) community
that is collectively facing it, including by exploiting the latest
achievements in Computer Science and adapting \emph{deep generative 
models} to parameterize the low-level response of the various 
experiments~\cite{Paganini:2017dwg, Krause:2021ilc, Amram:2023onf}. 
The literature refers to this kind of strategies with the term 
\emph{fast simulation}. Fast simulations share their data 
processing scheme and the reconstruction step with the detailed 
simulation (as depicted in Figure~\ref{fig:sim-proc-flow}), and 
are proven capable of reducing the computation cost of a simulated 
sample up to a factor of 20.

To meet the upcoming and future requests for simulated samples,
the LHCb Collaboration is also considering a more radical approach based
on the so-called \emph{ultra-fast simulation} paradigm. In this case,
the aim is to directly reproduce the high-level response of the
detector relying on a set of parameterizations developed to transform
generator-level particles information into reconstructed physics 
objects as schematically represented in Figure~\ref{fig:sim-proc-flow}
(bottom). Such parameterizations can still be built using generative
models, like \emph{Generative Adversarial Networks}~(GAN), proven 
to succeed in reproducing the high-level response of the LHCb 
detector~\cite{Ratnikov:2023wof} and offering reliable synthetic 
simulated samples~\cite{Anderlini:2022ckd}. Following pioneering 
studies on the ultra-fast simulation of the electromagnetic
calorimeter based on GANs~\cite{Musella:2018rdi}, the CMS Collaboration 
has recently started developing a full-scope ultra-fast simulation based on
\emph{Normalizing Flow}, named \flashsim~\cite{Vaselli:2858890}.

\section{\lamarr: the LHCb ultra-fast simulation framework}
\label{sec:lamarr}

\lamarr~\cite{Anderlini:2022ofl, Barbetti:2023bvi} is the official
ultra-fast simulation framework for LHCb, able to offer the fastest
options for simulation. Originating from the attempt of an LHCb 
customized version of \delphes~\cite{deFavereau:2013fsa, 
Siddi:2019abc}, \lamarr is an independent project retaining 
only the inspiration of its modular layout from \delphes.
In particular, the \lamarr framework consists of a pipeline of 
modular parameterizations, most of which based on machine learning 
algorithms, designed to take as input the particles generated by 
the physics generators and provide as output the high-level response 
of the various LHCb sub-detectors. 

The \lamarr pipeline can be logically split in two separated 
chains according to the charge of the generated particles. 
We expect that charged particles leave a mark in the Tracking 
system that \lamarr characterizes in terms of acceptance, 
efficiency and resolution as described in 
Section~\ref{sub:lamarr-track}. The reconstructed tracking 
variables are then used to compute the response of the PID 
system for a set of traversing charged particles (muons, pions, 
kaons or protons) as detailed in Section~\ref{sub:lamarr-pid}. 
In case of neutral particles (e.g., photons), the calorimeters 
play a key role and, since multiple photons can concur to the 
energy of a single calorimetric cluster, parameterizing 
particle-to-particle correlation effects is of major relevance. 
The solutions under investigation are reported in 
Section~\ref{sub:lamarr-calo}. The \lamarr pipelines described 
above are  shown in Figure~\ref{fig:lamarr-pipeline}.

\subsection{Tracking system}
\label{sub:lamarr-track}

\begin{figure}[h!]
  \begin{minipage}{0.57\textwidth}
    \centering
    \includegraphics[width=\textwidth]{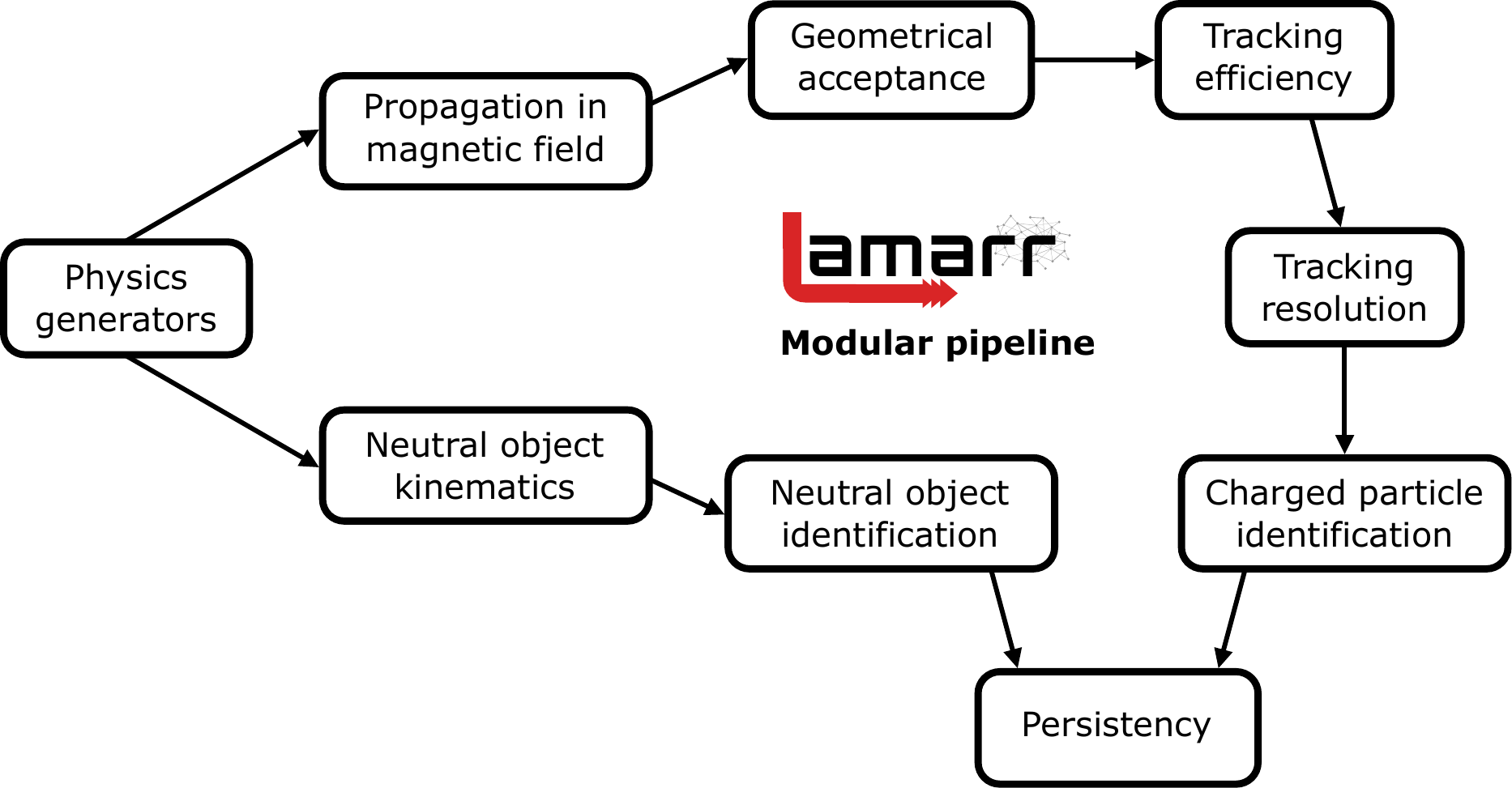}
  \end{minipage}
  \hfill
  \begin{minipage}{0.39\textwidth}
    \caption{
      \label{fig:lamarr-pipeline}
      Scheme of the \lamarr modular pipeline. According to the 
      charge of the particle provided by the physics generator, 
      two sets of parameterizations are defined: the charged 
      particles are passed through the Tracking and PID models, 
      while the neutral ones follow a different path where the 
      calorimeter modeling plays a key role.
    }
  \end{minipage}
\end{figure}

One of the aims of the LHCb Tracking system is to measure 
the momentum $p$ of charged particles (i.e., electrons, muons, 
pions, kaons and protons), exploiting the deflection of their 
trajectories due to the dipole magnet located in between the 
tracking detectors. Hence, the first 
step of the \emph{charged chain} reported in 
Figure~\ref{fig:lamarr-pipeline} is the propagation 
through the magnetic field of the particles provided by 
the physics generators. \lamarr parameterizes the 
particle trajectories as two rectilinear segments with a 
single deflection point (inversely proportional to the 
transverse momentum $p_T$), implementing the so-called 
\emph{single $p_T$ kick} approximation.

The next step requires to select the subset of tracks that fall
within the LHCb geometrical acceptance and that have any chance 
to be reconstructed. To this end, \lamarr
uses \emph{Gradient Boosted Decision Trees} (GBDT) trained to
learn the fraction of candidates that are in the acceptance as a
function of the kinematic information provided by the
physics generators. Given a generated track in acceptance, 
we ask whether the latter will be reconstructed and, in case
of positive answer, which tracking detectors are involved in
the reconstruction procedure. \lamarr statistically infers
such information, namely the tracking efficiency, relying 
on \emph{neural networks} trained to perform a multi-class 
classification according to the track kinematics. A major 
effort is ongoing to improve the performance of the efficiency
model on the basis of the type of tracks and particle species
(i.e., electrons, muons or hadrons).

At this point, \lamarr disposes of the subset of the generated 
particles that can be considered as reconstructed tracks,
but their kinematics and geometry are still identical to those 
provided by the physics generators. The smearing of these 
features, mimicking the effect of the reconstruction, is achieved 
using GANs. Driven by a \emph{binary cross-entropy} loss function 
and powered by \emph{skip connections}, GANs succeed in describing 
the resolution effects due to, for example, multiple scattering 
phenomena, only relying on track kinematic information 
at generator-level as input conditions. A similar GAN-based 
architecture is used to provide the correlation matrix obtained 
from the Kalman filter adopted in the reconstruction algorithm 
to define the position, slope and curvature of each track.

Stacking the parameterizations described above, \lamarr is able 
to provide the high-level response of the LHCb Tracking
system. The resulting reconstructed quantities can be further
processed using the LHCb analysis software to combine the 
parameterized tracks into decay candidates as depicted by 
the green slot in Figure~\ref{fig:sim-proc-flow} (bottom). 

\subsection{Particle identification system}
\label{sub:lamarr-pid}

To accomplish the LHCb physics program, disposing of 
a high-performance PID system is crucial since it allows for
discriminating the various particle species that traverse 
the detector. \lamarr provides parameterizations for the 
majority of the charged particles for which the PID
detectors are relevant (i.e., muons, pions, kaons
or protons). Specialized parameterizations for the electrons, 
encoding the multiple scattering and Bremsstrahlung emission 
contributions in the interaction with the detector materials, 
is planned as future development.

Identifying these subset particles involves mainly the RICH and 
MUON detectors, while the role played by the calorimeters is 
minor. In general, we expect that the response of the PID system
depends only on the specie of the traversing particle, its 
kinematics, and the detector occupancy. According to these
dependencies, \lamarr provides the high-level response for 
both the detectors using GAN-based models properly 
conditioned~\cite{Maevskiy:2019vwj, Anderlini:2022ckd}.
Given the particle specie from the physics generators, its
kinematic information results from the \lamarr Tracking modules, 
while the detector occupancy is described by the total 
number of tracks traversing the detector.

In real data, the combination of the responses from RICH
detectors, calorimeters, MUON system and a binary 
muon-identification criterion implemented via FPGA and 
named \texttt{isMuon} allows to compute the higher-level 
response of the PID system, referred to as GlobalPID variables.
The parameterization of the GlobalPID variables still relies
on conditioned GANs, adding as input what results from the
\texttt{RichGAN} and \texttt{MuonGAN} models. The binary output
of a neural-network-based implementation of \texttt{isMuon} is
used as additional input feature, while no explicit calorimeters
contribution is defined leaving the missing information problem
to the generator \emph{latent space}.

GAN-based models, driven by a \emph{Wasserstein distance} 
loss function and trained using a Lipschitz-constrained 
discriminator~\cite{Terjek:2020}, succeed in describing the
high-level response of the RICH and MUON systems. Chaining
together different GANs, \lamarr is also able to provide 
the higher-level response of the LHCb PID system, 
injecting an implicit contribution from the calorimeters.

\subsection{Electromagnetic calorimeter}
\label{sub:lamarr-calo}

Providing a parameterization for the electrons requires
describing the response to Bremsstrahlung photons by the 
LHCb ECAL detector. Since interested by a multitude of 
secondary particles, the detailed simulation of the 
calorimeter system is the most computationally expensive 
step in the simulation pipeline. The latter is a shared 
problem across the HEP community, that is investing 
great efforts in tuning deep generative models to 
properly parameterize the energy deposited in the calorimeter 
cells~\cite{Chekalina:2018hxi, Paganini:2017dwg, 
Krause:2021ilc, Amram:2023onf}. Such studies belong 
to the fast-simulation paradigm that aims to reduce 
the \geant use, providing models for the 
low-level response of the various experiments.

The current version of \lamarr provides a simplified 
parameterization for the LHCb calorimeter, designed for 
detector studies and based on a fast-simulation
approach. Disposing information at the calorimeter 
cell level requires running reconstruction algorithms to obtain
analysis-level quantities that may become rather CPU-expensive
for high-multiplicity events. In addition, since non-physical
strategies are used to simulate the energy deposits 
(as is the case for GANs), there is no certainty that 
the reconstruction software stack can correctly reproduce 
the expected distributions for the high-level 
variables~\cite{Rogachev:2022hjg}. Hence, 
the \lamarr project is actively working to provide an 
ultra-fast solution for the ECAL detector.

Reproducing the calorimeter high-level response is a 
non-trivial task since traditional generative models
rely on the hypothesis that an unambiguous relation between 
the generated particle and the reconstructed object 
exists\footnote{To a first approximation, the response of
the Tracking and PID systems satisfy this condition.}.
Instead, the presence of merged $\pi^0$ and Bremsstrahlung 
photons may lead to having $n$ generated particles responsible
for $m$ reconstructed objects (in general with $n \ne m$).
A strategy to face this particle-to-particle correlation
problem can be built using techniques designed in the context
of Language Modeling, describing the calorimeter simulation
as a \emph{translation problem}. To this end, 
\emph{Graph Neural Network}~(GNN)~\cite{Scarselli:2009abc} 
and \emph{Transformer}~\cite{Vaswani:2017abc} models are 
currently under investigation.

\begin{figure}[h!]
  \centering
  \includegraphics[width=\textwidth]{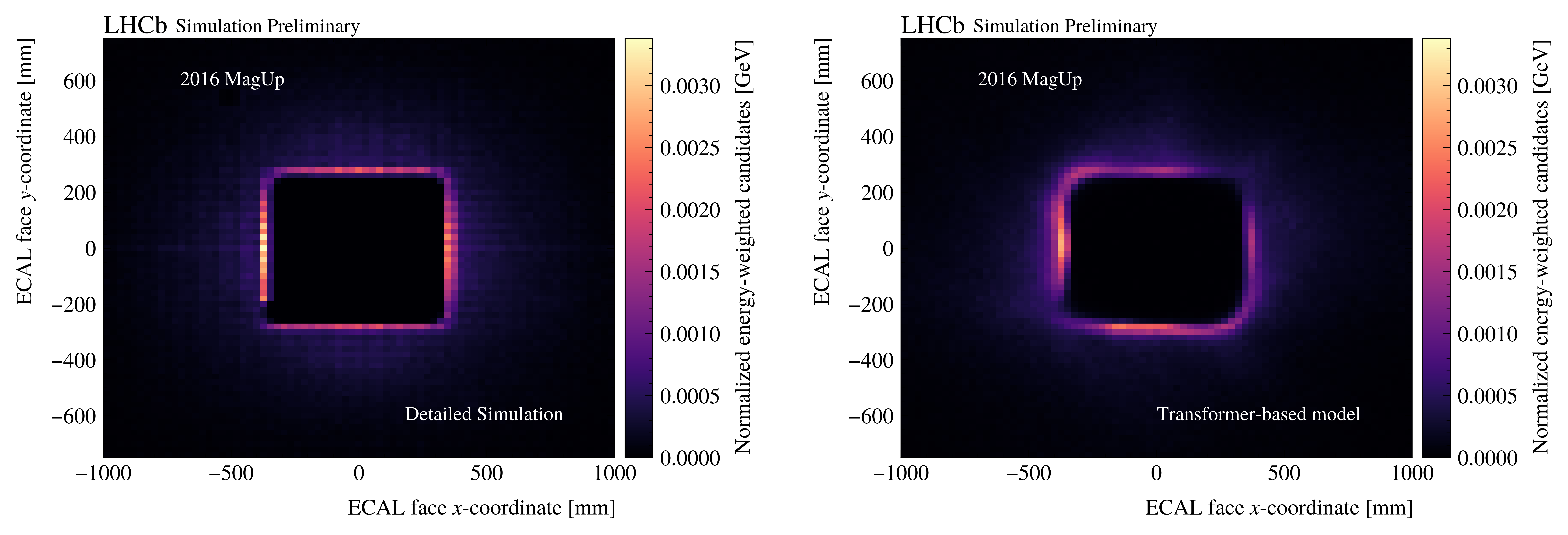}
  \caption{
    \label{fig:calotron-res}
    Distribution of the $(x, y)$-position of the reconstructed 
    clusters on the LHCb ECAL face for a $2000 \times 
    1500~\rm{mm}^2$ frame placed around the center. The 
    geometrical information is combined with the energy 
    signature properly weighting each bin entry. What 
    obtained from detailed simulation is reported on the left, 
    while the predictions of an adversarial trained Transformer
    model is shown on the right. The corresponding LHCB-FIGURE
    is in preparation.
  }
\end{figure}

Both the models are designed to process a sequence of
$n$ generated photons and infer the kinematics of a
sequence of $m$ reconstructed clusters. The non-trivial
correlations between any particles of the source
sequence (photons) and the target one (clusters) rely
on the \emph{attention mechanism}~\cite{Vaswani:2017abc, 
Brody:2022abc}. To improve the quality of the resulting 
parameterizations, the training of both GNN and 
Transformer-based models is driven by an adversarial
procedure (similarly to what occurs for GANs). The 
discriminator is currently implemented through a 
\emph{Deep Sets} model~\cite{Zaheer:2017abc}, while 
further studies are ongoing to replace it with a 
second Transformer~\cite{Lee:2022abc}. Considering the
complexity of the problem, the preliminary results
are promising as depicted in Figure~\ref{fig:calotron-res},
where the joint action of Transformer and Deep Sets
succeeds in deriving the energy distribution on
the ECAL face. The center of the calorimeter has not
active material since is used to host the LHC beam pipe.
It should be pointed out that no constraints are applied 
to the model output to reproduce such conditions, and that 
the empty space shown in Figure~\ref{fig:calotron-res} 
(right) is the result of the adversarial training procedure.

\section{Validation campaign and timing performance}
\label{sec:validation}

The ultra-fast philosophy at the base of the \lamarr framework
is being validated by comparing the distributions
obtained from machine-learnt models trained on detailed
simulation and the ones resulting from standard simulation
strategies. In particular, we will briefly discuss the 
validation studies performed for the charged particles pipeline
using simulated \LbLcmunu decays with \LcpKpi. The semileptonic
nature of the $\Lambda^0_b$ decay requires 
an interface with dedicated generators, in this case \evtgen.
Deeply studied by LHCb, this decay channel includes in its 
final state the four charged particle species parameterized 
in the current version of \lamarr, namely muons, pions, kaons 
and protons.

\begin{figure}[h!]
  \centering
  \includegraphics[width=0.49\textwidth]{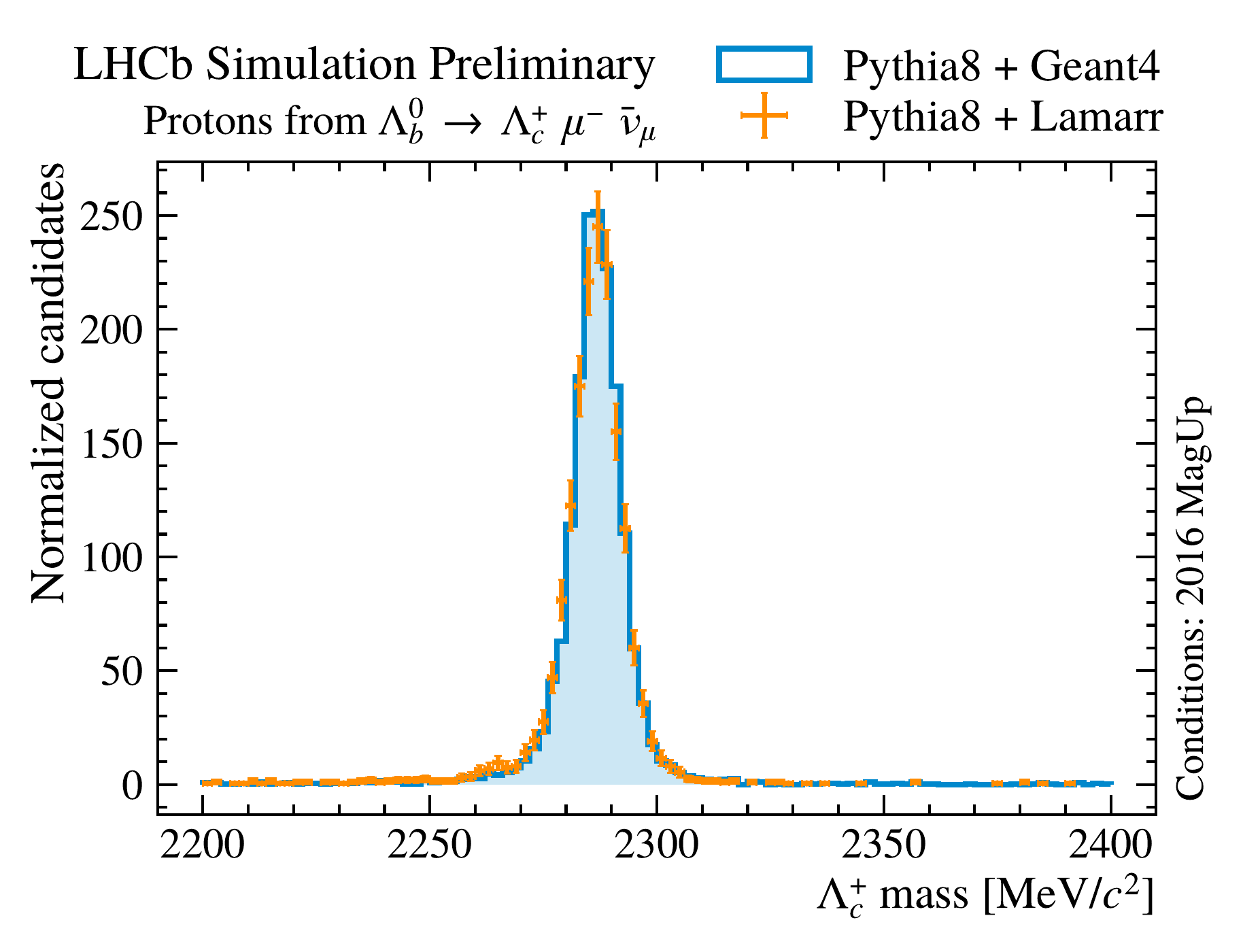}
  \hfill
  \includegraphics[width=0.49\textwidth]{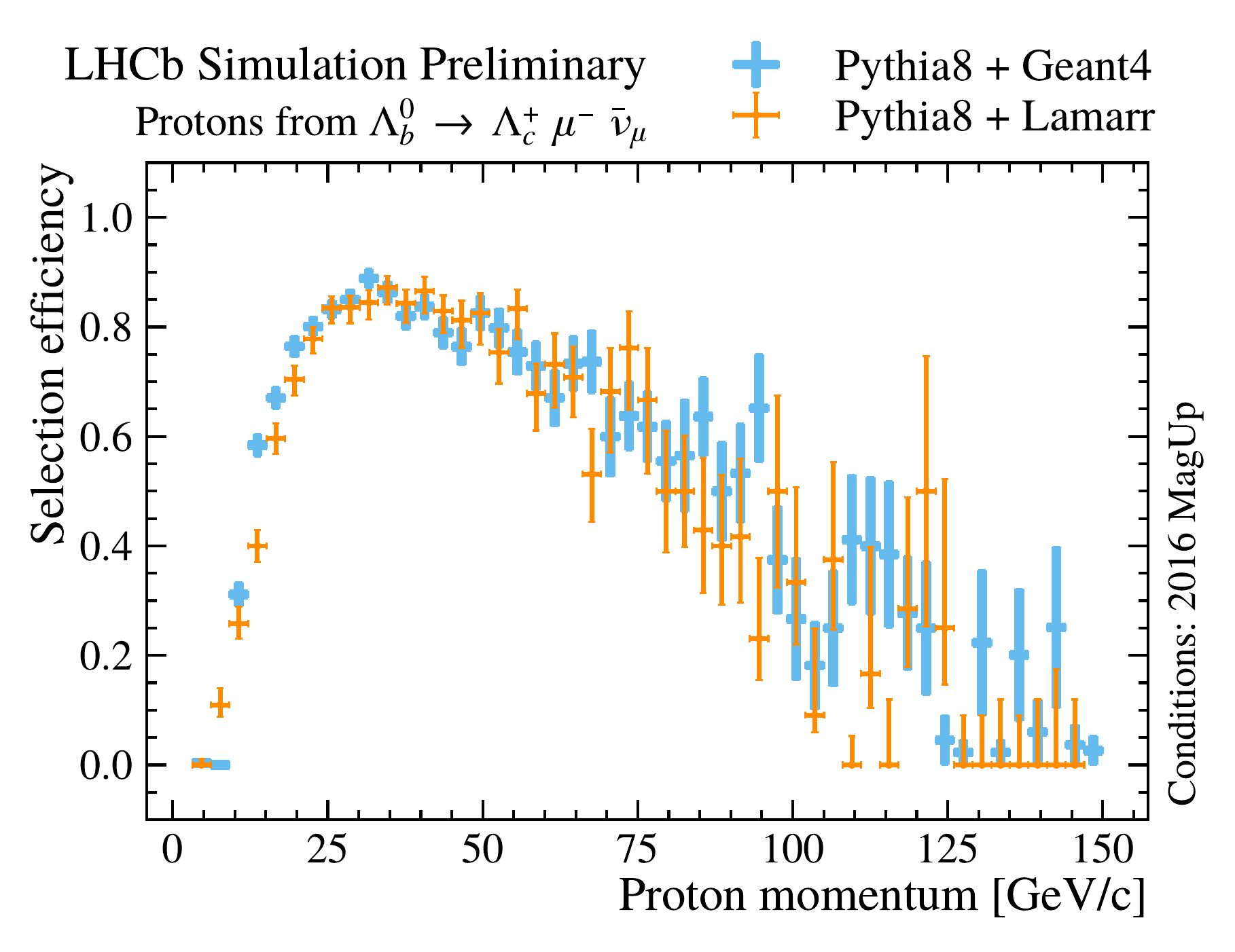}
  \caption{\label{fig:py8-res}
    Validation plots for \LbLcmunu decays with \LcpKpi 
    simulated with \pythia, \evtgen and \lamarr (orange 
    markers) and compared with detailed simulation samples 
    relying on \pythia, \evtgen and \geant (cyan shaded 
    histogram). Reproduced from 
    \href{https://cds.cern.ch/record/2814081}{LHCB-FIGURE-2022-014}.
  }
\end{figure}

The validation of the \lamarr Tracking modules is depicted in
Figure~\ref{fig:py8-res} (left) where the agreement between 
the $\Lambda^+_c$ invariant mass distribution resulting from
the ultra-fast paradigm and the one obtained from detailed 
simulation proves that the decay dynamics is well reproduced
and the resolution effects correctly parameterized. To show
the good performance of the \lamarr PID models, a comparison 
between the selection efficiencies for a tight requirement on 
a multivariate proton classifier is shown in
Figure~\ref{fig:py8-res} (right).

Comparing the CPU time spent per event by \geant-based
production of \LbLcmunu samples and the one needed by \lamarr,
we estimate a CPU reduction of two-order-of-magnitude only
for the simulation phase. Interestingly, since the generation
of $b$-baryons is exceptionally expensive, \pythia becomes the
major consumer of CPU resources in the ultra-fast paradigm.
A further speed-up can be reached reducing the cost for
generation, for example using a \emph{Particle Gun} that
simulates directly the signal particles without going 
through the high-energy collisions, not needed since
\lamarr parameterizes the detector occupancy. Even in 
these physics-simplified settings, the ultra-fast
philosophy succeeds in reproducing thee distributions 
obtained from detailed simulation~\cite{Anderlini:2022ofl}.

\section{Integration with the LHCb simulation framework}
\label{sec:integration}

To be integrated within the LHCb software stack, the parameterizations
provided by \lamarr need to be queried from a C++ application,
running in the \gaudi framework. Traditional deployment strategies
were found to lead to unacceptably large overheads due to the
presence of different multi-threading schedulers and context
switching issues. Hence, a custom deployment strategy was preferred:
models trained with \texttt{scikit-learn} and Keras are converted
into compatible C code using the \texttt{scikinC} 
toolkit~\cite{Anderlini:2022ltm}, and then distributed through
the LHCb Computing Grid via the CERN VM file-system 
(\texttt{cvmfs})~\cite{Buncic:2010zz}.

The modular layout of \lamarr enables a variety of studies and
developments on the single parameterizations, providing a
unique and shared infrastructure for validation and performance
measurements. While crucial for applications within LHCb, the
integration with \gaudi and \gauss makes the adoptions of \lamarr
unappealing for researchers outside of the LHCb community. The
SQLamarr package\footnote{Visit 
\url{https://lamarrsim.github.io/SQLamarr} for additional details.}
aims to mitigate this problem, providing a stand-alone ultra-fast
simulation framework with minimal dependencies. Based on SQLite3,
SQLamarr provides a set of classes and functions for loading data
from physics generators and defining pipelines from compiled models.
An integration between SQLamarr and \gaussino is currently under
investigation with the aim of providing ultra-fast parameterizations
following the experiment-independent philosophy of the newest LHCb
simulation framework, named \textsc{Gauss-on-Gaussino}\footnote{Visit 
\url{https://lhcb-gauss.docs.cern.ch/Futurev5} for additional 
details.}~\cite{Mazurek:2021abc, Mazurek:2022tlu}.

\section{Conclusion}
\label{sec:conclusion}

An evolution of the LHCb software stack and the simulation 
techniques are mandatory to meet the upcoming and future demand 
for simulated samples expected for \runIII and those that will follow.
Ultra-fast-based solutions will play a key role in 
reducing the pressure on pledged CPU resources, without
compromising unreasonably the description of the 
uncertainties introduced in the detection 
and reconstruction phases. Such techniques, powered by
deep generative models, are provided to LHCb via the novel 
\lamarr framework. Well integrated with the physics 
generators within the \gauss framework, \lamarr delivers two 
pipelines according to the charge of the generated particle. 
The statistical models for the Tracking and the charged PID 
systems have been deployed and validated with satisfactory 
results on \LbLcmunu decays. Several
models are currently under investigation for the neutral
pipeline, where the translation problem approach offers
a viable solution to face the particle-to-particle 
correlation problem. Further development of the integration
between \lamarr and the LHCb simulation framework is one
of the major ongoing activities to put the former in 
production and make its parameterizations available 
to the HEP community.

\section*{Acknowledgements}
This work is partially supported by ICSC -- Centro Nazionale 
di Ricerca in High Performance Computing, Big Data and Quantum 
Computing, funded by European Union -- NextGenerationEU.

\bibliography{main}

\begin{thebibliography}{31}

\bibitem{LHCb:2008vvz}
A.A. Alves, Jr. et~al. (LHCb), JINST \textbf{3}, S08005 (2008)

\bibitem{LHCb:2014set}
R.~Aaij et~al. (LHCb), Int. J. Mod. Phys. A \textbf{30}, 1530022 (2015), \texttt{1412.6352}

\bibitem{Clemencic:2011zza}
M.~Clemencic et~al. (LHCb), J. Phys. Conf. Ser. \textbf{331}, 032023 (2011)

\bibitem{Barrand:2001ny}
G.~Barrand et~al., Comput. Phys. Commun. \textbf{140}, 45 (2001)

\bibitem{Sjostrand:2007gs}
T.~Sjostrand, S.~Mrenna, P.Z. Skands, Comput. Phys. Commun. \textbf{178}, 852 (2008), \texttt{0710.3820}

\bibitem{Lange:2001uf}
D.J. Lange, Nucl. Instrum. Meth. A \textbf{462}, 152 (2001)

\bibitem{Allison:2006ve}
J.~Allison et~al., IEEE Trans. Nucl. Sci. \textbf{53}, 270 (2006)

\bibitem{Mazurek:2021abc}
M.~Mazurek, G.~Corti, D.~Müller, Comput. Inform. \textbf{40}, 815–832 (2021), \texttt{2112.04789}

\bibitem{Mazurek:2022tlu}
M.~Mazurek, M.~Clemencic, G.~Corti, PoS \textbf{ICHEP2022}, 225 (2023)

\bibitem{Chekalina:2018hxi}
V.~Chekalina et~al., EPJ Web Conf. \textbf{214}, 02034 (2019), \texttt{1812.01319}

\bibitem{Maevskiy:2019vwj}
A.~Maevskiy et~al. (LHCb), J. Phys. Conf. Ser. \textbf{1525}, 012097 (2020), \texttt{1905.11825}

\bibitem{Anderlini:2022ofl}
L.~Anderlini et~al., PoS \textbf{ICHEP2022}, 233 (2023)

\bibitem{Barbetti:2023bvi}
M.~Barbetti (2023), {ACAT'22}, \texttt{2303.11428}

\bibitem{Paganini:2017dwg}
M.~Paganini, L.~de~Oliveira, B.~Nachman, Phys. Rev. D \textbf{97}, 014021 (2018), \texttt{1712.10321}

\bibitem{Krause:2021ilc}
C.~Krause, D.~Shih, Phys. Rev. D \textbf{107}, 113003 (2023), \texttt{2106.05285}

\bibitem{Amram:2023onf}
O.~Amram, K.~Pedro (2023), \texttt{2308.03876}

\bibitem{Ratnikov:2023wof}
F.~Ratnikov et~al., Nucl. Instrum. Meth. A \textbf{1046}, 167591 (2023)

\bibitem{Anderlini:2022ckd}
L.~Anderlini et~al. (LHCb), J. Phys. Conf. Ser. \textbf{2438}, 012130 (2023), \texttt{2204.09947}

\bibitem{Musella:2018rdi}
P.~Musella, F.~Pandolfi, Comput. Softw. Big Sci. \textbf{2}, 8 (2018), \texttt{1805.00850}

\bibitem{Vaselli:2858890}
F.~Vaselli et~al. (CMS), Tech. rep. (2023), \urlstyle{tt}\url{https://cds.cern.ch/record/2858890}

\bibitem{deFavereau:2013fsa}
J.~de~Favereau et~al. (DELPHES 3), JHEP \textbf{02}, 057 (2014), \texttt{1307.6346}

\bibitem{Siddi:2019abc}
B.G. Siddi, EPJ Web Conf. \textbf{214}, 02024 (2019)

\bibitem{Terjek:2020}
D.~Terjék (2020), {ICLR'20}, \texttt{1907.05681}

\bibitem{Rogachev:2022hjg}
A.~Rogachev, F.~Ratnikov, J. Phys. Conf. Ser. \textbf{2438}, 012086 (2023), \texttt{2207.06329}

\bibitem{Scarselli:2009abc}
F.~Scarselli et~al., IEEE Trans. Neural. Netw. \textbf{20}, 61 (2009)

\bibitem{Vaswani:2017abc}
A.~Vaswani et~al. (2017), {NeurIPS'17}, \texttt{1706.03762}

\bibitem{Brody:2022abc}
S.~Brody, U.~Alon, E.~Yahav (2022), {ICLR'22}, \texttt{2105.14491}

\bibitem{Zaheer:2017abc}
M.~Zaheer et~al. (2017), {NeurIPS'17}, \texttt{1703.06114}

\bibitem{Lee:2022abc}
K.~Lee et~al. (2022), {ICLR'22}, \texttt{2107.04589}

\bibitem{Anderlini:2022ltm}
L.~Anderlini, M.~Barbetti, PoS \textbf{CompTools2021}, 034 (2022)

\bibitem{Buncic:2010zz}
P.~Buncic et~al., J. Phys. Conf. Ser. \textbf{219}, 042003 (2010)

\end{thebibliography}

\end{document}